\documentstyle[12pt]{article}   
\begin{document}
\begin{center}   
\Large
{\bf{Linear Systems with adiabatic fluctuations 
\footnote{Dedicated to Prof. Mihir Chowdhury
on his 60th birthday}
}}
\end{center}

\vspace{0.2cm}

\begin{center}
{\bf{Suman Kumar Banik and Deb Shankar Ray}}

{\bf{Indian Association for the Cultivation of Science}}

{\bf{Jadavpur , Calcutta 700 032 , INDIA.}}

\end{center}

\vspace{3.0cm}

\begin{abstract}
We consider a dynamical system subjected to weak but adiabatically slow
fluctuations of external origin. Based on the ``adiabatic following'' 
approximation we carry out an expansion in $\alpha|\mu|^{-1}$, where 
$\alpha$ is the strength of fluctuations and $|\mu|^{-1}$ refers to the time
scale of evolution of the unperturbed system to obtain a linear differential 
equation for the average solution. The theory is applied to the problems of
a damped harmonic oscillator and diffusion in a turbulent fluid. The result 
is the realization of `renormalised' diffusion constant or damping constant
for the respective problems. The applicability of the method has been
critically analyzed.
\end{abstract}

\vspace{3.0cm}

PACS NO.  05.20 Dd , 05.40  +j

\newpage

%%%%%%%%%%%%%%%%%%%%%%%%%%%%%%%%%%%%%%%%%%%%%%%%%%%%%%%%%%%%%%%%%%%%%%%%%%%%%

\begin{center}
\large{
{\bf{I.\hspace{0.2cm}Introduction}}}
\end{center}

\vspace{0.5cm}  

The standard paradigm of the temporal evolution of nonequilibrium processes
regarded, in general, as stochastic processes is the century-old problem of
Brownian motion [1,2]. This involves the random motion of microscopic particles
effectively introducing the motion of a physical system, the Brownian 
particle to be observed on a macroscopic level. To generate the successive
levels of description from the microscopic to the macroscopic realm one
essentially introduces coarse-graining of space and time in the dynamics.
Although there exists no general program of coarse-graining, it is
nevertheless possible to realize the dynamics of stochastic processes in terms
of some systematic separation of time scales consistent with the experiments
at the macroscopic level of description.

\vspace{0.2cm}  

The standard separation of time scales in the description of Brownian motion
involves correlation functions which are nonzero over some interval
$\tau_{c}$ which is the correlation time of fluctuations and we require
that $\Delta t$, the coarse-grained timescale over which one observes the 
average motion is much greater than $\tau_{c}$, such that
$\gamma^{-1}\gg \Delta t \gg \tau_{c}$, where $\gamma^{-1}$
is the system's damping time. Physically this implies that one smoothes out the
fluctuations of the system on a time scale during which microscopic particles
are correlated but not on a scale during which the system is damped. Thus
the fluctuations considered in the stochastic process of the Brownian
motion are weak and rapid.

\vspace{0.2cm}

In the present problem we consider a multivariate dynamical system driven by
weak but
adiabatically slow fluctuations. The slow fluctuations characterized by very 
long correlation time have also attracted a lot of attention of various workers
over the years [2-4,7,8,11,14,15]. While the overwhelming majority of the
treatment of stochastic differential equation with fast fluctuations is based
on the assumption that there is a very short auto-correlation time $\tau_{c}$,
such that one can adopt the scheme of expansion in $\alpha \tau_{c}$, 
suitable simplifying approximation for dealing with very long correlation-time
is relatively scarce. In general, the problem of long correlation time is
theoretically handled at the expense of severe restriction on the type of
stochastic behavior. For instance, several authors [2-4,7,8,11,14,15]
have tried the linear and nonlinear models within the framework of Markov processes
of the type dichotomic processes, two-state Markov processes, random telegraphic
processes etc. Our aim here is to explore a {\it {perturvative method}} for finding an
equation for the average solution pertaining to the separation 
of timescale implied in the inequality

\begin{eqnarray*}
\frac{1}{|\mu|} \ll \Delta t \ll \tau_{c} \hspace{0.2cm},
\end{eqnarray*}

\noindent
$|\mu|$ being the largest eigenvalue of the unperturbed system, {\it{where we do
not keep any restriction on the type of stochastic behavior}}. The strategy
being perturbative is based on an expansion in $\alpha |\mu|^{-1}$ rather
than in $\alpha \tau_{c}$ as is done in the case of fast fluctuations. The 
method dealt in the present treatment is thus somewhat complementary to the 
scheme of expansion of the latter kind.

\vspace{0.2cm}

To put the issue in a proper perspective we first borrow a simple example 
of adiabatic dynamics in terms of Bloch equations [5,6], wellknown in 
magnetic resonance and quantum optical experiments. The problem concerns
a two-level system interacting with a single mode electromagnetic field,
where the field ${\cal E}(t)$ varies slowly enough ``adiabatically'' in 
the time scale of the inverse of the damping constant or frequency
detuning between the atom and the field. The term ``adiabatic
following'' is thus used to describe collectively the associated
experimental phenomena [5,6]. The model is described by the following 
equation,

\vspace{0.2cm}

\begin{equation}
\frac{d}{d t}\left ( \begin{array}{c}
u\\
v\\
w \end{array}\right )=\left ( \begin{array}{ccc}
-\frac{1}{T_{2}} & -\Delta & 0\\
\Delta & -\frac{1}{T_{2}} & g{\cal E}(t)\\
0 & -g{\cal E}(t) & -\frac{1}{T_{1}} \end{array}\right )
\left ( \begin{array}{c}
u\\
v\\
w \end{array}\right ) + \left ( \begin{array}{c}
0\\
0\\
\frac{w_{\rm eq}}{T_{1}} \end{array}\right ) \hspace{0.5cm} .
\end{equation}

\vspace{0.2cm}

Here $u,v,w$ are the Bloch vector components, $T_{1}$ and $T_{2}$  are the
energy and dephasing relaxation terms, $\Delta$ is the detuning of the 
frequency of the field ${\cal E}(t)$ from that of the two-level system. $g$
includes the effect of coupling of the atom to the field. The equilibrium
value, towards which the population inversion $w$ relaxes when
${\cal E}=0$ is denoted by $w_{\rm eq}$. Adiabatic
following approximation asserts that if the field ${\cal E}(t)$ is varied
slowly enough then $w$, the population inversion variable would follow
adiabatically from $-1$ to $\sim +1$ in the process, i.e., a ground state
population could be adiabatically inverted.

\vspace{0.2cm}

Our problem in the present investigation concerns such processes where the 
adiabatic variation of ${\cal E}(t)$, in addition, is stochastic. Thus the
usual limit in the ``adiabatic following'' applies, i.e., the rate of 
variation of the pulse or fluctuations is much small compared to the
relaxation rate of the system. With these in mind we may treat Eq.(1) as a
stochastic differential equation provided the stochastic properties of
${\cal E}(t)$ are a priori known. 

\vspace{0.2cm}   

To formulate the problem we thus consider a system subject to fluctuating 
external forces where the fluctuations are weak and adiabatically slow. The
equation of motion then become a stochastic differential equation, a particular
category of which is a general form of Eq.(1) $[$For simplicity we disregard
the constant part on the right hand side$]$,

\vspace{0.2cm}

\begin{equation}
{\dot{u}}={\bf{A}}(t) \hspace{0.05cm} u\hspace{0.05cm} .
\end{equation}

\vspace{0.2cm}

Here ${\bf{A}}(t)$ is a random function of time, stochastic properties of 
which are given. Linear multiplicative noise ( Eq.(2) ) has got wide
application in studying the random Markov process [7], fluctuating barrier
crossing [8], enzymatic kinetics in biology [9], nuclear magnetic resonance
in physics [10] and stochastic resonance in linear system [11] and in many 
other context [12].

\vspace{0.2cm}

Based on the systematic separation of time scales using adiabatic following
approximation a differential equation for the average solution $\langle 
u \rangle$ is obtained. This approximation allows us an expansion in 
$\alpha | \mu|^{-1}$, where $\alpha$ is a measure of the strength of 
fluctuations and $\mu^{-1}$ refers to the internal time scale of the unperturbed 
system.

\vspace{0.2cm}

As an immediate application of the method we treat the problem of a damped
harmonic oscillator with adiabatically fluctuating frequency. The method is
extended to the problem of diffusion in a turbulent fluid as another 
illustration. The central result is that one realizes a ``renormalised'' 
transport coefficient or a damping constant so that the diffusion or the 
damping process gets significantly modified by adiabatic stochasticity. We 
show that the method is equipped to deal with similar kinds of stochastic 
processes.

\vspace{0.2cm}

The outlay of the paper is as follows; In Sec.II we discuss the method of
adiabatic following approximation on the stochastic differential equation
of the form (2). The essential idea is to extract the average dynamics of
the relevant physical quantity. The method has been critically analyzed
in Sec.III. The method is applied to two specific cases in Sec.IV. 
We point out that a wide class of problems can be treated in a similar way. 
The paper is concluded in Sec.V.

%%%%%%%%%%%%%%%%%%%%%%%%%%%%%%%%%%%%%%%%%%%%%%%%%%%%%%%%%%%%%%%%%%%%%%%%%%%%%

\vspace{0.5cm}

\begin{center}
\large{
{\bf{II. \hspace{0.2cm} A method for weak and adiabatic fluctuations}}}
\end{center}

\vspace{0.5cm}

To start with we consider a linear equation of the type (2) and rewrite
it as

\vspace{0.2cm}

\begin{equation}
{\dot{u}}=\{ {\bf{A}}_{0} + \alpha {\bf{A}}_{1}(t)\} u
\hspace{0.05cm},
\end{equation}

\vspace{0.2cm}

\noindent
where $u$ is a vector with $n$ components, ${\bf{A}}_{0}$ is a constant 
matrix of dimension $n\times n$ and ${\bf{A}}_{1}(t)$ is a stochastic matrix, 
$\alpha$ is a parameter ( of dimension $1/t$ ) which measures the strength of 
fluctuation.

\vspace{0.2cm}

It is convenient to assume that ${\bf{A}}_{1}(t)$ is a stationary process 
with $\langle {\bf{A}}_{1}(t)\rangle=0$. Eq.(3) admits of two important 
time scales of the system measured by the inverse of the largest eigenvalue
of the matrix ${\bf{A}}_{0}$  and the time scale of fluctuations 
of ${\bf{A}}_{1}(t)$ (the correlation time of fluctuation). As already been 
mentioned that in the treatment of overwhelming majority of the stochastic 
processes, such as, motion of a Brownian particle in a fluid or electromagnetic 
waves in a turbulent atmosphere, one essentially considers a situation
where the fluctuations are weak and rapid. 
The correlation time of fluctuations is much short compared 
to the time scale set by the inverse of the eigenvalues of ${\bf{A}}_{0}$. In 
the appropriate limit we encounter the delta-correlated events and solve 
approximately or exactly the relevant stochastic differential equations [2]. The 
familiar examples of paramagnetic resonance and line broadening are well known 
in this context.

\vspace{0.2cm}

Since in the present problem we consider a stochastic process in which the 
fluctuations are weak but adiabatically slow, ${\bf{A}}_{1}(t)$ is  an 
adiabatic stochastic process. Therefore the usual procedure of systematic 
cumulant expansion which inherently takes into account of short correlation
time of fluctuations is not valid. An alternative treatment is thus sought
for. 

\vspace{0.2cm}
To this end we first introduce an interaction representation as given
by,

\vspace{0.2cm}

\hspace{2.5cm}$ u(t)=\exp({\bf{A}}_{0} t) v(t)$

\vspace{0.5cm}

\noindent
and applying it to Eq.(3) we obtain,

\vspace{0.5cm}

\hspace{2.5cm}${\dot{v}}=\alpha {\bf{V}}(t) v$ \hspace{0.05cm},

\vspace{0.5cm}

\noindent
where, \hspace{1.3cm}${\bf{V}}(t)=\exp(-{\bf{A}}_{0} t) {\bf{A}}_{1}(t)\exp({\bf{A}}
_{0} t)$ \hspace{0.05cm}.

\vspace{0.5cm}

On integration the last equation yields,

\vspace{0.2cm}

\begin{equation}
v(t)=v(0)+\alpha\int_{0}^{t} {\bf{V}}(t')v(t') dt'
\hspace{0.05cm} .
\end{equation}

\vspace{0.2cm}

On iterating the Eq.(4) once, we are now led to an ensemble average equation 
of the following form,

\vspace{0.2cm}

\begin{equation} 
\langle v(t)\rangle=v(0)+\alpha^{2}\int_{0}^{t} dt'\int_{0} 
^{t'} d t'' \langle {\bf{V}}(t'){\bf{V}}(t'') v(t'')\rangle \hspace{0.05cm} .
\end{equation} 

\vspace{0.2cm}

The equation is still exact since no second order approximation (as usually
done) has been used.

\vspace{0.2cm}

Now taking the time derivative of $v(t)$ we arrive at the following
integrodifferential equation in which the initial value $v(0)$ no
longer appears,

\vspace{0.2cm}

\begin{equation} 
\frac{d}{d t}\langle v(t)\rangle=\alpha^{2}\int_{0}^{t}  
\langle {\bf{V}}(t){\bf{V}}(t') v(t')\rangle dt' \hspace{0.05cm}.
\end{equation} 

\vspace{0.2cm}

Making use of a change of integration variable $t'=t-\tau$ we obtain,

\vspace{0.2cm}

\begin{equation} 
\frac{d}{d t}\langle v(t)\rangle=\alpha^{2}\int_{0}^{t}  
\langle {\bf{V}}(t){\bf{V}}(t-\tau) v(t-\tau)\rangle d\tau \hspace{0.05cm}.
\end{equation} 

\vspace{0.2cm}

Reverting back to the original representation Eq.(7) yields

\vspace{0.2cm}

\begin{equation} 
\frac{d}{d t}\langle u(t)\rangle={\bf{A}}_{0}\langle u\rangle
+\alpha^{2}\int_{0}^{t}  
\langle {\bf{A}}_{1}(t)\exp({\bf{A}}_{0}\tau){\bf{A}}_{1}(t-\tau)
u(t-\tau) \rangle d\tau \hspace{0.05cm} .
\end{equation} 

\vspace{0.2cm}

The adiabatic following assumption, that ${\bf{A}}_{1}(t)$ and the components
of $u(t)$ vary slowly on the scale of inverse of ${\bf{A}}_{0}$, can 
now be utilized. Following Crisp[6] we note that a Taylor series expansion of
${\bf{A}}_{1}(t-\tau) u(t-\tau)$ in the average $\langle\ldots\rangle$
of the $\alpha^{2}$-term in Eq.(8) allows the integral to be evaluated
and the last equation reduces to the following form,

\vspace{0.2cm}

\begin{equation} 
\frac{d}{d t}\langle u(t)\rangle={\bf{A}}_{0}\langle u\rangle
+\alpha^{2}\sum_{n=o}^{\infty}\frac{(-1)^{n}}{n !}  
\langle {\bf{A}}_{1}(t) \left\{ \int_{0}^{\infty} d\tau\exp({\bf{A}}_{0}
\tau)\tau^{n} \right\}\frac{d^{n}}{d t^{n}}[{\bf{A}}_{1}(t) u(t)] 
\rangle \hspace{0.5cm} . 
\end{equation} 

\vspace{0.2cm}

The integral in Eq.(9) can be evaluated by rewriting it in terms of the 
following matrix elements

\vspace{0.2cm}

\begin{eqnarray*}
I_{ik}^{n}
=\int_{0}^{\infty} d\tau\tau^{n}\sum_{j}
D_{ij}e^{\mu_{jj}\tau}D_{jk}^{-1} \hspace{0.5cm} ,
\end{eqnarray*}

\begin{eqnarray*}
\hspace{3.0cm}=\sum_{j} D_{ij}\frac{n!}{(\mu_{jj})^{n+1}}D_{jk}^{-1},
\hspace{1.0cm} {\rm Re}\hspace{0.2cm} \mu_{jj} < 0 \hspace{0.5cm} .
\end{eqnarray*} 

\vspace{0.2cm}

Therefore,

\vspace{0.2cm}

\begin{equation}
{\bf{I}}^{n}=n!\hspace{0.1cm}{\bf{D}}\hspace{0.1cm}{\bf{E}}_{n+1}
\hspace{0.1cm}{\bf{D}}^{-1} \hspace{0.05cm},
\end{equation}

\vspace{0.2cm}

\noindent
where ${\bf{D}}$ is a matrix which diagonalises ${\bf{A}}_{0}$ and

\vspace{0.2cm}

\begin{eqnarray*}
{\bf{E}}_{n+1}=\left(\begin{array}{ccc}
\frac{1}{\mu_{11}^{n+1}} & & 0\\
& \ddots &\\
0 & & \frac{1}{\mu_{jj}^{n+1}}\end{array}\right ) 
\end{eqnarray*}

\vspace{0.2cm}

\noindent
and $\mu_{jj}$ are the eigenvalues of ${\bf{A}}_{0}$.

\vspace{0.2cm}

The Eq.(9) then assumes the following form,

\vspace{0.2cm}

\begin{equation} 
\frac{d}{d t}\langle u(t)\rangle={\bf{A}}_{0}\langle u\rangle
+\alpha^{2}\sum_{n=o}^{\infty}(-1)^{n}  
\langle {\bf{A}}_{1}(t) {\bf{D}} {\bf{E}}_{n+1}{\bf{D}}^{-1}
\frac{d^{n}}{d t^{n}}[ {\bf{A}}_{1}(t) u(t) ] \rangle \hspace{0.05cm}.
\end{equation} 

\vspace{0.2cm}

Although the equation involves an infinite series it is expected to yield
useful result in the adiabatic following approximation. If this 
approximation is valid, the quantity $[{\bf{A}}_{1}(t) u(t)]$ varies
little in the time $|\mu_{jj}^{n+1}|^{-1}$  of ${\bf{E}}_{n+1}$ and
the series converges rapidly. Keeping only the two lowest order terms we
arrive at,

\vspace{0.2cm}

\begin{eqnarray} 
\frac{d}{d t}\langle u(t)\rangle={\bf{A}}_{0}\langle u\rangle
+\alpha^{2}\langle {\bf{A}}_{1}(t){\bf{X}}_{1}{\bf{A}}_{1}(t) u(t)
\rangle -\alpha^{2}\langle {\bf{A}}_{1}(t){\bf{X}}_{2}{\dot{\bf{A}}_{1}}
(t) u(t)\rangle\nonumber\\
-\alpha^{2}\langle {\bf{A}}_{1}(t){\bf{X}}_{2}{\bf{A}}_{1}(t){\dot{u}}(t)
\rangle
\end{eqnarray} 

\vspace{0.2cm}

\noindent
where,

\vspace{0.2cm}

\begin{equation}
{\bf{X}}_{n+1}={\bf{D}}\hspace{0.1cm}{\bf{E}}_{n+1}\hspace{0.1cm}
{\bf{D}}^{-1} \hspace{0.05cm}.
\end{equation}

\vspace{0.2cm}

It is evident that the average $\langle {\dot{u}} \rangle$ is related to 
a more complicated average. As a next approximation [13-15] 
we now suppose that the latter averages may be broken up as,

\vspace{0.2cm}

\begin{equation} 
\langle {\bf{A}}_{1}(t){\bf{X}}_{1}{\bf{A}}_{1}(t) u(t)\rangle \approx
\langle {\bf{A}}_{1}(t){\bf{X}}_{1}{\bf{A}}_{1}(t)\rangle
\langle u(t)\rangle
\end{equation} 

\vspace{0.2cm}

\noindent
and so on. Keeping only the terms upto the order of $\alpha^{2}$ we obtain,

\vspace{0.2cm}    

\begin{eqnarray} 
\frac{d}{dt}\langle u(t)\rangle=\left \{ {\bf{A}}_{0}
+\alpha^{2}[ \langle {\bf{A}}_{1}(t){\bf{X}}_{1}{\bf{A}}_{1}(t)\rangle-
\langle {\bf{A}}_{1}(t){\bf{X}}_{2}{\dot{\bf{A}}_{1}}(t)\rangle
\right. \nonumber\\
\left.-\langle {\bf{A}}_{1}(t){\bf{X}}_{2}{\bf{A}}_{1}(t)\rangle{\bf{A}}_{0}]
\right\} \langle u(t)\rangle \hspace{0.05cm} .
\end{eqnarray} 

\vspace{0.2cm}    

Thus the average of $u(t)$ obeys a nonstochastic differential equation
in which the effect of weak adiabatic fluctuations is accounted for by
``renormalizing'' ${\bf{A}}_{0}$ through the addition of constant terms
of the order of $\alpha^{2}$. The net effect is that depending on the
specificity of the situations one realizes a dissipative or a gain term in 
the average dynamics. We note in passing, that the average dynamics of
$u$ is independent of any explicit correlation function.

\vspace{0.5cm}

\begin{center}
\large{
\bf{III.\hspace{0.2cm}Discussions on the method}}
\end{center}

\vspace{0.5cm}

Theory of stochastic differential equations with multiplicative noise has
a long history. The stochastic processes dealt with the overwhelming 
majority of the cases concern the fast processes (more precisely, the 
correlation time between the noise events has been considered to be the 
shortest timescale of the dynamics). In the previous section we have 
considered a stochastic process which adiabatically slow. The traditional 
scheme of solving stochastic differential equations with fast noise processes
is that one reduces them to Bourret's integral equations [13] and then 
performs the decoupling of the product of operators. In the present paper
we have followed equation scheme upto Eq.(9) and then make
use of ``adiabatic following'' approximation. It is necessary to make the 
following distinctions;

\vspace{0.2cm}

First, note that in going from Eq.(8) to (11) we have made no approximation
so far as the full infinite series is concerned. Also each term is not of order 
$\alpha\tau_{c}$ as in the case of fast processes (as emphasized by van Kampen
[14]) but of order $\alpha\frac{d^{n}}{dt^{n}}[{\bf{A}}(t)u(t)]/\mu_{jj}
^{n+1}$. Just as the theory of fast processes is valid for $\alpha\tau_{c}$
very small which implies that the successive cumulants in the expansion are
small, validity of the description of adiabatic processes rests on the
smallness of successive $\alpha\frac{d^{n}}{dt^{n}}[{\bf{A}}(t)u(t)]/\mu_{jj}
^{n+1}$ terms. Thus the two expansions are essentially different.

\vspace{0.2cm}

Second, the decoupling approximation has been carried out both in the fast as 
well as in the slow processes. Its justifications in the former case has been 
established  early by Brissaud and Frisch [15]. It had been strongly advocated 
by van Kampen [14] who has asserted that although it seems to neglect certain
correlations, the ``statistical mechanics of transport processes would be in 
a very sorry state without such hypothesis''. It is not difficult to 
comprehend that its spiritual root lies in ``stosszahlansatz'', ``molecular 
chaos assumption'' or ``random phase approximations''. The essential point, 
however, in the decoupling scheme is the realization of a separation of time
scale of average of the product of fluctuating quantities ${\bf{A}}(t)$ and 
the average of $u$ itself consistent with the expansions pertaining to slow 
or fast processes.

\vspace{0.2cm}

That the two expansion schemes in the fast and slow stochastic processes
are different can be confirmed if one compares the lowest order terms of 
the corresponding evolution. According to the present scheme Eq.(15) itself 
asserts that (free motion neglected)

\vspace{0.2cm}

\begin{equation}
\frac{d}{d t}\langle u \rangle \sim \frac{\alpha^{2}}{|\mu|}\langle u\rangle
\hspace{0.2cm},
\end{equation}

\vspace{0.2cm}

\noindent
where $|\mu|^{-1}$ refers to the timescale set by the ${\bf{A}}_{0}$ matrix
which is short in the adiabatic following limit. For a fast stochastic
process on the other hand the counterpart of Eq.(16) is

\vspace{0.2cm}

\begin{equation}
\frac{d}{d t}\langle u \rangle \sim \alpha^{2}\tau_{c}\langle u\rangle
\hspace{0.2cm},
\end{equation}

\vspace{0.2cm}

\noindent
where $\tau_{c}$ defines the very short correlation time of the noise [14].

\vspace{0.2cm}

The difference in the expansion schemes also makes the relative errors made 
in the decoupling approximation in the two cases, different. To this end we 
first note that Eq.(15) is obtained from Eq.(8). Upto second order it means 
omitting terms of the order $(\alpha\Delta t)^{3}$ and higher (where $\Delta 
t$ defines the coarse-grained time scale of evolution of the average). As the
lower bound of $\Delta t$ is determined by $|\mu|^{-1}$, it implies that we
neglect terms of the order $(\alpha |\mu|^{-1})^{3}$ in the evolution equation. 
Thus the relative error made in the decoupling approximation in the case of 
adiabatic expansion is $(\alpha |\mu|^{-1})^{3}$.

\vspace{0.2cm}

As demonstrated by van Kampen [14] the corresponding error made in the decoupling 
approximation in Bourret's scheme is of the order $(\alpha\tau_{c})^{3}$. The
workability of the decoupling approximation in the fast and slow stochastic 
processes is thus demonstrated in the two different expansion procedures
ensuring their respective fast convergence in the limit $\alpha\tau_{c}$ 
(fast processes) or $\alpha |\mu|^{-1}$ (slow processes) small but finite.

\vspace{0.2cm}

So, to summarize, we point out that the
implementation of Bourret's decoupling approximation is a major step 
for almost any treatment of multiplicative noise upto date [2,7,11-15]. This
is because of the fact that the average of a product of stochastic quantities 
does not factorize into the product of averages, although it has been argued 
that [2,7,11-21] good approximations can be derived by assuming such factorization.
In the case of fast fluctuations it has been justified if the driving stochastic
noise has a fast correlation time such that Kubo number $\alpha^2\tau_{c}$ is
very small in the cumulant expansion scheme ( an expansion in $\alpha\tau_{c}$
). The factorization has been shown to be exact in the limit of zero 
correlation time and in some specific noise processes [7,14] and the solution 
for the average can be found exactly. The present scheme of adiabatic expansion on the other hand
is an expansion in $\alpha |\mu|^{-1}$ and it may be argued in the same way
that factorization in the slow fluctuation is valid where
$\alpha^2 |\mu|^{-1}$ is very small. Essentially it implies that $u(t)$
in the average ( in the right hand side of Eq.(12) ) is realized as an
average $\langle u(t)\rangle$ ( which varies in the coarse-grained timescale
$\Delta t$ ) pertaining to the separation of the timescales in the inequality 
$|\mu|^{-1} \ll \Delta t \ll \tau_{c} $ adopted in the present case instead
of the inequality $\tau_{c} \ll \Delta t \ll |\mu|^{-1}$ employed in the case 
of fast fluctuation and cumulant expansion.

\vspace{0.5cm}

\begin{center}
\large{
\bf{IV.\hspace{0.2cm}Applications}}
\end{center}

\vspace{0.2cm}

\begin{center}
\bf{A.\hspace{0.2cm}Damped harmonic oscillator with adiabatically fluctuating
frequency}
\end{center}

\vspace{0.3cm}

To illustrate the above-mentioned method we consider a model of damped
harmonic oscillator with random frequency where the fluctuation is weak and
adiabatically slow in the time scale of dissipation. The opposite limit of 
weak and rapid fluctuations in frequency has been studied by numerous 
authors in connection with turbulence, wave propagation, line-broadening [10],
lasers, chaotic dynamics [20,21]. A comprehensive treatment has been given
in van Kampen [14].

\vspace{0.2cm}

We are now in a position to apply the result(15) to the following equation,

\vspace{0.2cm}

\begin{equation}
\ddot{x}+\omega^{2}(t)x+\gamma \dot{x}=0 \hspace{0.5cm} ,
\end{equation}

\vspace{0.2cm}

\noindent
with an adiabatically stochastic frequency,

\vspace{0.2cm}

\begin{equation}
\omega^{2}(t)=\omega_{0}^{2}[1+\alpha\xi(t)] \hspace{0.05cm} ,
\end{equation}

\vspace{0.2cm}

\noindent
where $\xi(t)$ is an adiabatic stochastic process with zero mean $\langle
\xi(t)\rangle=0$; $\omega_{0}$ is the frequency of the unperturbed 
oscillator and $\gamma$ is the damping constant. $\alpha$, the smallness 
parameter is dimensionless in Eq.(19).

\vspace{0.2cm}

Rewriting Eq.(18) in the form,

\vspace{0.2cm}

\begin{equation}
\frac{d}{dt}\left ( \begin{array}{c}
x\\
y \end{array} \right )=\left ( \begin{array}{cc}
0 & 1\\
-\omega_{0}^{2} & -\gamma \end{array} \right ) \left ( \begin{array}{c}
x\\
y \end{array}\right )+\alpha\omega_{0}^{2}\xi (t)\left ( \begin{array}{cc}
0 & 0\\
-1 & 0 \end{array}\right )\left ( \begin{array}{c}
x\\
y \end{array}\right )
\end{equation}

\vspace{0.2cm}

\noindent
one identifies,

\vspace{0.2cm}

\begin{equation}
{\bf{A}}_{0}=\left ( \begin{array}{cc}
0 & 1\\
-\omega_{0}^{2} & -\gamma \end{array} \right ) ;
{\bf{A}}_{1}=\omega_{0}^{2}\xi (t)\left ( \begin{array}{cc}
0 & 0\\
-1 & 0 \end{array} \right ) ;
u(t)=\left ( \begin{array}{c}
x\\
y \end{array} \right )
\end{equation}

\vspace{0.2cm}

\noindent
of Eq.(15). ${\bf{X}}_{1}$ and ${\bf{X}}_{2}$ are related to ${\bf{E}}_{1}$ 
and ${\bf{E}}_{2}$ through (13) and are given by,

\vspace{0.2cm}

\begin{eqnarray*}
\left.{\bf{X}}_{1}=\left ( \begin{array}{cc}
\frac{B-A}{AB} & \frac{1}{AB}\\
\\
1 & 0 \end{array} \right ) \right.;
\left.{\bf{X}}_{2}=\left ( \begin{array}{cc}
\frac{A^{2}-AB+B^{2}}{(AB)^{2}} & \frac{B-A}{(AB)^{2}}\\
\\
\frac{B-A}{AB} & \frac{1}{AB}\end{array} \right ) \right.
\end{eqnarray*}

\vspace{0.05cm} 

\noindent
and

\vspace{0.05cm} 

\begin{eqnarray*}
\left.{\bf{E}}_{1}=\left ( \begin{array}{cc}
\frac{1}{A} & 0\\
\\
0 & -\frac{1}{B}\end{array} \right ) \right.;
\left.{\bf{E}}_{2}=\left ( \begin{array}{cc}
\frac{1}{A^{2}} & 0\\
\\
0 & \frac{1}{B^{2}}\end{array} \right ) \right. \hspace{0.05cm},
\end{eqnarray*}

\vspace{0.2cm}

\noindent
where $A$ and $B$ are related to the eigenvalues $(e_{1},e_{2})$ of
${\bf{A_{0}}}$ matrix :

\vspace{0.2cm}

\begin{eqnarray*}
A=-\frac{\gamma}{2}+\frac{1}{2}\sqrt{\gamma^{2}-4\omega_{0}^{2}},
\hspace{0.2cm} (e_{1}) \hspace{0.05cm},\\
B=\frac{\gamma}{2}+\frac{1}{2}\sqrt{\gamma^{2}-4\omega_{0}^{2}},
\hspace{0.2cm} (-e_{2}) \hspace{0.05cm}.
\end{eqnarray*}

\vspace{0.2cm}
{\bf{D}} matrix is given by ,

\vspace{0.2cm}

\begin{eqnarray*}
{\bf{D}}=\left ( \begin{array}{cc}
\frac{1}{({A+1})^{\frac{1}{2}}} & \frac{1}{({B+1})^{\frac{1}{2}}}\\ 
\\
\frac{A}{({A+1})^{\frac{1}{2}}} & \frac{-B}{({B+1})^{\frac{1}{2}}} \end{array}
\right ) \hspace{0.05cm}.
\end{eqnarray*}

\vspace{0.2cm}

So for the present problem, the Eq.(15) takes the form,

\vspace{0.2cm}

\begin{eqnarray*}
\frac{d}{dt}\left ( \begin{array}{c}
\langle x \rangle\\
\langle y \rangle \end{array} \right )=\left ( \begin{array}{cc}
0 & 1\\
-\omega_{0}^{2}-\alpha^{2}\omega_{0}^{2}\langle\xi^{2}\rangle-\alpha^{2}
\gamma\langle\xi\dot{\xi}\rangle & -\gamma+\alpha^{2}\langle\xi^{2} 
\rangle\gamma\end{array} \right ) \left ( \begin{array}{c}
\langle x \rangle\\
\langle y \rangle \end{array} \right ) \hspace{0.05cm},
\end{eqnarray*}

\vspace{0.2cm}

\noindent
or

\vspace{0.2cm}

\begin{equation}
\langle\ddot{x}\rangle+\gamma[1-\alpha^{2}\langle\xi^{2}\rangle
]\langle\dot{x}\rangle+[\omega_{0}^{2}+\alpha^{2}\omega_{0}^{2}\langle
\xi^{2}\rangle+\alpha^{2}\gamma\langle\xi\dot{\xi}\rangle]\langle x\rangle
=0 \hspace{0.05cm}.
\end{equation}

\vspace{0.2cm}

It is thus evident that the adiabatic fluctuations in frequency cause a
suppression of damping of the average amplitude of the oscillator. Or, in
other words, the dissipative oscillator experiences a partial gain in average 
amplitude by an amount,

\vspace{0.2cm}

\begin{equation}
\gamma_{\rm gain}=\alpha^{2}\gamma\langle\xi^{2}(t)\rangle \hspace{0.05cm}.
\end{equation}

\vspace{0.2cm}

As expected, the frequency of the unperturbed oscillator has also undergone
a shift in addition to this gain in amplitude.

\vspace{0.2cm}

The above result can be compared to the case of fast fluctuations in  
frequency as treated by van Kampen and others. It is interesting to note 
that while the adiabatic fluctuations result in gain in amplitude the fast
fluctuations cause a damping of the average amplitude, in general. This 
damping may even be negative when the fluctuations are particularly strong 
at twice the unperturbed frequency. The latter results had been found to be
useful in the context of a fluctuation-dissipation relation in chaotic
dynamics [21].

\vspace{0.2cm}

The theory developed in the preceding section also permits us to calculate
the dynamics of the higher moments. For example, the equations of the three 
moments can be found from Eq.(18),

\vspace{0.2cm}

\begin{equation}
\frac{d}{dt}\left ( \begin{array}{c}
x^{2}\\
xy\\
y^{2} \end{array} \right )=\left ( \begin{array}{ccc}
0 & 2 & 0\\
-\omega^{2} & -\gamma & 1\\
0 & -2\omega^{2} & -2\gamma \end{array}\right )\left (\begin{array}{c}
x^{2}\\
xy\\
y^{2} \end{array} \right ) \hspace{0.05cm},
\end{equation}

\vspace{0.2cm}

\noindent
or rewriting it in the form,

\vspace{0.2cm}

\begin{equation}
\frac{d}{dt}\left ( \begin{array}{c}
x^{2}\\
xy\\
y^{2} \end{array} \right )=\left ( \begin{array}{ccc}
0 & 2 & 0\\
-\omega_{0}^{2} & -\gamma & 1\\
0 & -2\omega_{0}^{2} & -2\gamma \end{array}\right )\left (\begin{array}{c}
x^{2}\\
xy\\
y^{2} \end{array} \right )+\alpha\omega_{0}^{2}\xi(t)
\left ( \begin{array}{ccc}
0 & 0 & 0\\
-1 & 0 & 0\\
0 & -2 & 0 \end{array}\right )\left (\begin{array}{c}
x^{2}\\
xy\\
y^{2} \end{array} \right ) \hspace{0.05cm},
\end{equation}

\vspace{0.2cm}

\noindent
we identify,

\vspace{0.2cm} 

\begin{eqnarray*}
{\bf{A}}_{0}=\left ( \begin{array}{ccc}
0 & 2 & 0\\
-\omega_{0}^{2} & -\gamma & 1\\
0 & -2\omega_{0}^{2} & -2\gamma \end{array}\right )\hspace{0.2cm}
{\rm and}\hspace{0.2cm}
{\bf{A}}_{1}(t)=\alpha\omega_{0}^{2}\xi(t)\left ( \begin{array}{ccc}
0 & 0 & 0\\
-1 & 0 & 0\\
0 & -2 & 0 \end{array}\right )\hspace{0.2cm}.
\end{eqnarray*}

The eigenvalues of ${\bf{A}}_{0}$ are,

\begin{eqnarray}
\left.\begin{array}{lll}
e_{1}=-\gamma \\
e_{2}=-\gamma+\sqrt{\gamma^{2}-4\omega_{0}^{2}} \\
e_{3}=-\gamma-\sqrt{\gamma^{2}-4\omega_{0}^{2}}
\end{array} \right \} \hspace{0.05cm}.
\end{eqnarray}

\vspace{0.2cm}

Eq.(15) therefore takes the form of an evolution equation of higher moments,

\vspace{0.2cm}

\begin{equation}
\frac{d}{dt}\left (\begin{array}{c}
\langle x^{2}\rangle\\
\\
\langle xy \rangle\\
\\
\langle y^{2}\rangle \end{array} \right )={\bf{T}}\hspace{0.2cm}
\left (\begin{array}{c}
\langle x^{2}\rangle\\
\\
\langle xy \rangle\\
\\
\langle y^{2}\rangle \end{array} \right)\hspace{0.2cm}
\end{equation}

\vspace{0.2cm}

\noindent
where

\begin{equation}
{\bf{T}}=
\left (\begin{array}{ccc}
0 & 2 & 0\\
\\
\frac{\alpha^{2}\omega_{0}^{2}(\gamma^{2}c_{1}-\gamma c_{2}+2 \omega_{0}^{2}
c_{1})}{2\gamma^{2}}-\omega_{0}^{2} & \frac{\alpha^{2}(\gamma^{3}c_{1}-
2\omega_{0}^{2}\gamma c_{1}-2 \omega_{0}^{2} c_{2}-\gamma^{2} c_{2})}
{2\gamma^{2}}-\gamma & 1-\frac
{\alpha^{2}(\gamma^{2}+2 \omega_{0}^{2}) c_{1}}{2\gamma^{2}}\\
\\
\frac{\alpha^{2}\omega_{0}^{2}(\gamma c_{2}-\gamma^{2} c_{1}-\omega_{0}^{2}
c_{1})}{\gamma} & \frac{\alpha^{2}(\gamma^{2}c_{2}+
\omega_{0}^{2} c_{2}-\gamma^{3} c_{1}+3\omega_{0}^{2}\gamma c_{1})}
{\gamma}-2 \omega_{0}^{2} & \frac{\alpha^{2}(\gamma^{2}+\omega_{0}^{2}) c_{1}}
{\gamma}-2\gamma
\end{array}\right )
\end{equation}

\vspace{0.2cm}

\noindent
where $c_{1}=\langle\xi^{2}\rangle$ and $c_{2}=\langle\xi\dot{\xi}\rangle
\hspace{0.2cm}.$

\vspace{0.2cm}

What follows as a consequence of Eq.(27) is the shifting of eigenvalues
of the unperturbed oscillator. The eigenvalue which corresponds to $e_{1}$
of the unperturbed case now becomes,

\vspace{0.2cm}

\begin{eqnarray*}
-\gamma-\alpha^{2}\left \{\frac{\gamma}{2}c_{1}+\frac{ \omega_{0}^{2}}
{4\omega_{0}^{2}-\gamma^{2}}c_{1}-\frac{1}{2}c_{2}\right\}
\end{eqnarray*}

\vspace{0.2cm}

\noindent
to second order in $\alpha$. Thus the damping of energy of the unperturbed
oscillator gets enhanced beyond a critical value depending on the positivity
of the term included in the parenthesis of the last expression. In the 
negative region the term acts as a gain term leading to a suppression of 
dissipation of energy of the oscillator. 

\vspace{0.2cm}

In contrast to this case of adiabatic fluctuations, fast fluctuations make 
the oscillator unstable energy-wise due to the fluctuations in the force  
that have twice the frequency of the oscillator. Addition of damping term,
however, may result in stability below a certain critical value and 
unstability above it. This result has been particularly relevant in 
establishing the Kubo relation in chaotic dynamics [20].

\vspace{0.2cm}

\begin{center}
\bf{B.\hspace{0.2cm} Diffusion in a turbulent fluid}
\end{center}

\vspace{0.5cm}

As a next application we consider a diffusive process in a fluid in motion
described by,

\vspace{0.2cm}

\begin{equation}
\frac{\partial n}{\partial t}=-\nabla\cdot (n{\bf{v}})+D\nabla^{2} n 
\hspace{0.05cm}.
\end{equation}

\vspace{0.2cm}

Here $n({\bf{r}},t)$ is the number of `probe particles' per unit volume and
${\bf{v}}({\bf{r}},t)$ is the velocity of the moving fluid. If turbulence sets in
then ${\bf{v}}({\bf{r}},t)$ becomes a stochastic function of ${\bf{r}}$ and $t$.
The problem has been addressed by numerous workers over several decades.
To quote a representative few of them we refer to Ref [14].

\vspace{0.2cm} 

For the present problem we consider ${\bf{v}}({\bf{r}},t)$ as an adiabatic 
stochastic process. The problem then is to find an average $n({\bf{r}},t)$ 
for the given initial condition

\vspace{0.2cm} 

\begin{eqnarray*}
n({\bf{r}},0)=\delta({\bf{r}}) \hspace{0.05cm}.
\end{eqnarray*}

\vspace{0.2cm} 

We consider the turbulence to be weak and slow and without any loss of
generality assume

\vspace{0.2cm} 

\begin{eqnarray*}
\langle{\bf{v}}({\bf{r}},t)\rangle=0
\end{eqnarray*}

\vspace{0.2cm} 

Eq.(29) is of the form (3), provided the matrix ${\bf{A}}_{0}$ and 
${\bf{A}}_{1}$ correspond to

\vspace{0.2cm}

\begin{eqnarray*}
{\bf{A}}_{0}=D\nabla^{2} \hspace{0.2cm} {\rm and} \hspace{0.2cm}
\alpha{\bf{A}}_{1}=\nabla\cdot{\bf{v}}
\end{eqnarray*}

\vspace{0.2cm}

The symbol $\nabla$, as usual, acts on every functions of ${\bf{r}}$ that
appears to the right of it. Eq.(8) then takes the following form,

\vspace{0.2cm}

\begin{equation}
\frac{\partial}{\partial t}\langle n({\bf{r}},t)\rangle=D\nabla^{2}
\langle n({\bf{r}},t)\rangle+\int_{0}^{t} d\tau\langle \nabla \cdot
{\bf{v}}({\bf{r}},t) e^{\tau D\nabla^{2}}\nabla\cdot{\bf{v}}({\bf{r}},t-\tau)
n({\bf{r}},t-\tau)\rangle \hspace{0.05cm}.
\end{equation}

\vspace{0.2cm}

We take the Fourier transform in space of the last Eq.(30) to obtain,

\vspace{0.2cm}

\begin{eqnarray}
\frac{\partial}{\partial t}\langle n({\bf{k}},t)\rangle=-D k^{2}
\langle n({\bf{k}},t)\rangle-(2\pi)^{-3}\sum_{i,j}\int_{0}^{t} d\tau
\int_{-\infty}^{\infty}\int_{-\infty}^{\infty}\langle 
v_{i}({\bf{q}},t) v_{j}({\bf{q}}',t-\tau)\nonumber\\ 
n({\bf{k}},t-\tau)\rangle
\times (q_{i}+q_{i}'+k_{i}) e^{-\tau D({\bf{k}}+{\bf{q}}')^{2}}
(q_{j}'+k_{j}) d{\bf{q}} d{\bf{q}}'\hspace{0.05cm}.
\end{eqnarray}

\vspace{0.2cm}

Expanding $v_{j}({\bf{q}}',t-\tau)n({\bf{k}},t-\tau)$ as a Taylor series
and integrating over $\tau$  as before we arrive at,

\vspace{0.2cm}

\begin{eqnarray}
\frac{\partial}{\partial t}\langle n({\bf{k}},t)\rangle=-D k^{2}
\langle n({\bf{k}},t)\rangle-(2\pi)^{-3}\sum_{i,j}\sum_{n}(-1)^{n}
\int_{-\infty}^{\infty}\int_{-\infty}^{\infty}\langle 
v_{i}({\bf{q}},t)\nonumber\\ 
\frac{1}{[D({\bf{k}}+{\bf{q}}')^{2}]^{n+1}} 
\partial_{t}^{n}[v_{j}({\bf{q}}',t) n({\bf{k}},t)]\rangle
\times (q_{i}+q_{i}'+k_{i}) 
(q_{j}'+k_{j}) d{\bf{q}} d{\bf{q}}' \hspace{0.05cm}.
\end{eqnarray}

\vspace{0.2cm}

Taking into account of the following properties of the stochastic function
${\bf{v}}({\bf{q}},t)$,

\vspace{0.2cm}

\begin{eqnarray}
\langle v_{i}({\bf{q}},t)v_{j}({\bf{q}}',t)\rangle=\delta({\bf{q}}+{\bf{q}}')
(2\pi)^{\frac{3}{2}} C_{ij}({\bf{q}}) \hspace{0.05cm},\nonumber\\
\langle v_{i}({\bf{q}},t)\dot{v}_{j}({\bf{q}}',t)\rangle=\delta({\bf{q}}+
{\bf{q}}') (2\pi)^{\frac{3}{2}} C_{ij}'({\bf{q}}) \hspace{0.05cm},
\end{eqnarray}

\vspace{0.2cm}

and imposing the adiabatic following approximation that $v_{j}({\bf{q}},t-
\tau)n({\bf{k}},t-\tau)$ varies much slowly in the time scale of $D^{-1}$
we obtain,

\vspace{0.2cm}

\begin{eqnarray}
\frac{\partial}{\partial t}\langle n({\bf{k}},t)\rangle=\left \{-Dk^{2}-
(2\pi)^{-\frac{3}{2}}\sum_{i,j}\int_{-\infty}^{\infty} C_{ij}({\bf{q}})k_{i}
(k_{j}-q_{j})\frac{1}{D({\bf{k}}-{\bf{q}})^{2}} d{\bf{q}}\right.\nonumber\\
\left. +(2\pi)^{-\frac{3}{2}}\sum_{i,j}\int_{-\infty}^{\infty} C_{ij}'
({\bf{q}})k_{i}(k_{j}-q_{j})\frac{1}{D^{2}({\bf{k}}-{\bf{q}})^{4}} 
d{\bf{q}}\right.\nonumber\\
\left.-(2\pi)^{-\frac{3}{2}}\sum_{i,j}\int_{-\infty}^{\infty} C_{ij}
({\bf{q}})k_{i} (k_{j}-q_{j})\frac{k^{2}}{D({\bf{k}}-{\bf{q}})^{4}} d{\bf{q}}
\right \} \hspace{0.05cm}.
\end{eqnarray}

\vspace{0.2cm}

The effect of adiabatic stochasticity in the motion of the fluid thus 
essentially is to recover a renormalised diffusion coefficient $D(k)$ of
`test' particles in the following form,

\vspace{0.2cm}  

\begin{eqnarray}
D(k)=D+
(2\pi)^{-\frac{3}{2}}\sum_{i,j}\int_{-\infty}^{\infty} C_{ij}({\bf{q}})k_{i}
(k_{j}-q_{j})\frac{1}{Dk^{2}({\bf{k}}-{\bf{q}})^{2}} d{\bf{q}}\nonumber\\
-(2\pi)^{-\frac{3}{2}}\sum_{i,j}\int_{-\infty}^{\infty} C_{ij}'
({\bf{q}})k_{i}(k_{j}-q_{j})\frac{1}{D^{2}k^{2}({\bf{k}}-{\bf{q}})^{4}} 
d{\bf{q}}\nonumber\\
+(2\pi)^{-\frac{3}{2}}\sum_{i,j}\int_{-\infty}^{\infty} C_{ij}
({\bf{q}})k_{i} (k_{j}-q_{j})\frac{1}{D({\bf{k}}-{\bf{q}})^{4}} d{\bf{q}}
\hspace{0.05cm}.
\end{eqnarray}

\vspace{0.2cm}  

\noindent
Hence Eq.(34) reduces to,

\vspace{0.2cm}  

\begin{equation}
\frac{\partial}{\partial t}\langle n({\bf{k}},t)\rangle=-D(k) k^{2}
\langle n({\bf{k}},t)\rangle
\end{equation}

\vspace{0.2cm}

\noindent
a normalized diffusion equation for isotropic turbulence. In the absence of
any detailed knowledge about the stochastic properties embedded in 
$C_{ij}({\bf{q}})$ and $C_{ij}'({\bf{q}})$ it is difficult to proceed 
further. Nevertheless in the limit of small ${\bf{q}}$ one might expect
some interesting behavior as has been observed in the case of rapid
fluctuations.

\vspace{0.2cm}

It is well known that the long wavelength fast fluctuations are insufficiently
damped by the viscosity, (which appears as a parameter in the correlation
function of the incompressible fluids) which ensures the existence of a 
finite $\tau_{c}$. This causes long time tails in the correlation functions.
As van Kampen [14] emphasized the stochastic description in terms of an average
$\langle n \rangle$ ceases to become meaningful in these cases. Since the 
present treatment of slow fluctuations is free from explicit correlation 
functions and we deal only with averages, such pathological problems of long 
time tails or memory need not trouble us to that extent. This leads us to 
believe that the average description remains more meaningful in such cases.

\vspace{0.5cm}

\begin{center}
\large{
\bf{V.\hspace{0.2cm} Conclusion}}
\end{center}

\vspace{0.5cm}

In this paper we develop a method for treatment of weak but adiabatically
slow stochastic process. Based on the ``adiabatic following'' approximation we
recast a class of linear stochastic differential equations with multiplicative
noise into a differential equation for the average solution. This has been
carried on the basis of an expansion in $\alpha|\mu|^{-1}$, where $\alpha$
is the size of the fluctuation and $|\mu|^{-1}$ refers to the time scale of
evolution of the unperturbed system. The result differs significantly from
the corresponding treatment of weak and rapid fluctuations which relies on the
expansion in $\alpha\tau_{c}$, where $\tau_{c}$ is the auto-correlation time
of fluctuations [14]. It is also necessary to emphasize that in the 
present work no a priori assumption on the nature of noise in ${\bf{A}}_{1}
(t)$ (like ${\bf{A}}_{1}(t)$ is a Gaussian random process etc., which has 
received so much attention in the literature) has been made. Although in our
applications we have dealt with classical and linear problems, the theory
can be extended to quantum mechanical and nonlinear problems as well. We hope
to address such issues in future communications.

\newpage

{\bf{Acknowledgments}} : Thanks are due to Dr.B.Deb ( Dept. of Physical 
Chemistry, I.A.C.S. ) and Prof.N.C.Sil ( Dept. of Theoretical Physics, 
I.A.C.S. ) for interesting discussions.
The partial financial support from DST (Govt. of India) is also thankfully
acknowledged.

\newpage

{\bf{References}} :

\begin{enumerate}
\item Selected Papers on Noise and Stochastic Processes, edited by N. Wax
( Dover Publ. New york, 1954 ).
\item Stochastic Processes in Physics and Chemistry, N. G. van Kampen (North
Holland Physics Pub., New york, 1981 ).
\item R. C. Bourret, U. Frisch and A. Pouquet, Physica {\bf 65}, 303 (1973) ;
O. J. Heilman and N. G. Van Kampen, Physica {\bf 77}, 279 (1974) ; M. R.
Cruty and K. C. So, Phys. Fluids. {\bf 16}, 1765 (1973).
\item S. Faetti, L. Fronzoni and P. Grigolini, Phys. Rev. {\bf A32}, 1150
(1985) ; N. Nakasako and M. Ohta, Acoust. Letts. {\bf 8}, 199 (1985) ;
P. Schramm and I. Oppenheim, Physica {\bf 137A}, 81 (1986).
\item D. Grischkowsky, Phys. Rev. Letts. {\bf{24}}, 866 (1970) ; 
D. Grischkowsky, E. Courtens and J. A. Armstrong , Phys. Rev. Letts.
{\bf{31}}, 422 (1973). 
\item M. D. Crisp, Phys. Rev. {\bf{A8}}, 2128 (1973).
\item H. F. Arnoldus and G. Nienhuis, J. Phys. A {\bf 19}, 1629 (1986).
\item A. Fulinski, Phys. Lett. A {\bf 180}, 94 (1993) ; A. J. R. Madureira,
P. H$\ddot{\rm a}$nggi, V. Buonomano and W. A. Rodrigues, Phys. Rev. {\bf E51},
3849 (1995) and the references therein.
\item B. Robertson and R. D. Astumian, Bioph. J. {\bf 91}, 4891 (1991) ;
J. Chem. Phys. {\bf 94}, 7414 (1991).
\item R. Kubo, J. Math. Phys. {\bf{4}}, 174 (1963).
\item V. Berdichevsky and M. Gitterman, Europhys. Lett. {\bf 36}, 161 (1996).
\item M. Rahman, Phys. Rev. {\bf E52}, 2486 (1995) ; H. K. Leung and B. C. Lai,
Phys. Rev. {\bf E47}, 3043 (1993) ; R. Walsen, H. Ritsch, P. Zoller and J.
Cooper, Phys. Rev. {\bf A45}, 468 (1992).
\item R. C. Bourret, Can. J. Phys. {\bf{40}}, 782 (1962) ; 
Nuovo. Cim. {\bf{26}}, 1 (1962).
\item N. G. van Kampen, Phys. Reps. {\bf{24}}, 171 (1976).
\item A. Brissaud and U. Frisch, J. Math. Phys. {\bf{15}}, 524 (1974).
\item R. F. Fox, J. Math. Phys. {\bf 13}, 1196 (1972).
\item H. F. Arnoldus and G. Nienhuis, J. Phys. B {\bf 16}, 2325 (1983).
\item Z. Deng and J. H. Eberly, Opt. Commun. {\bf 51}, 189 (1984).
\item V. E. Shapiro and V. M. Loginov, Physica {\bf 91A}, 563 (1978).
\item S. Chaudhuri, G. Gangopadhyay and D. S. Ray, Phys. Rev. {\bf{E47}}, 311 
(1993). 
\item S. Chaudhuri, G. Gangopadhyay and D. S. Ray, Phys. Rev. {\bf{E52}}, 
2262 (1995). 
\end{enumerate}
\end{document}